\documentclass[9pt,twocolumn,twoside]{pnas-new}

\usepackage[utf8]{inputenc}
\usepackage[T1]{fontenc}

\usepackage{amsmath}
\usepackage{amssymb}
\usepackage{amsfonts}
\usepackage{pifont}
\usepackage{graphicx}
\usepackage{siunitx}
\usepackage{xcolor}
\usepackage{physics}

\DeclareMathAlphabet\mathbfcal{OMS}{cmsy}{b}{n}

\newcommand{\x}{\mathbf{x}}
\newcommand{\X}{\mathbf{X}}
\newcommand{\K}{K}
\newcommand{\bK}{\mathbf{K}}
\newcommand{\W}{\mathcal{W}}
\newcommand{\bW}{\mathbfcal{W}}
\newcommand{\Q}{\mathcal{Q}}

\newcommand{\V}{\mathcal{V}}

\newcommand{\methods}{\emph{Material and Methods}}

\newcommand*\patchAmsMathEnvironmentForLineno[1]{%
 \expandafter\let\csname old#1\expandafter\endcsname\csname #1\endcsname
 \expandafter\let\csname oldend#1\expandafter\endcsname\csname end#1\endcsname
 \renewenvironment{#1}%
 {\linenomath\csname old#1\endcsname}%
 {\csname oldend#1\endcsname\endlinenomath}}%
\newcommand*\patchBothAmsMathEnvironmentsForLineno[1]{%
 \patchAmsMathEnvironmentForLineno{#1}%
 \patchAmsMathEnvironmentForLineno{#1*}}%
\AtBeginDocument{%
\patchBothAmsMathEnvironmentsForLineno{equation}%
\patchBothAmsMathEnvironmentsForLineno{align}%
\patchBothAmsMathEnvironmentsForLineno{flalign}%
\patchBothAmsMathEnvironmentsForLineno{alignat}%
\patchBothAmsMathEnvironmentsForLineno{gather}%
\patchBothAmsMathEnvironmentsForLineno{multline}%
}

\templatetype{pnasresearcharticle}

\title{Adiabatic computing for optimal thermodynamic efficiency of information processing}

\author{Salamb\^{o} Dago}
\author{Sergio Ciliberto}
\author[1]{Ludovic Bellon}

\affil{Univ Lyon, ENS de Lyon, CNRS, Laboratoire de Physique, F-69342 Lyon, France}

\leadauthor{Dago} 

\significancestatement{To face the climate transition challenges, frugal computing is a rising trend towards energy efficient devices. At the crossroad between statistical physics and information theory, stochastic thermodynamics can provide fundamental tools and guidelines for sustainable applications. We study here a basic logical operation that is known to come with an incompressible energetic cost: a 1-bit memory erasure. Lowering the dissipation of the physical system realizing the memory seems desirable, but comes at the expense of a higher thermal insulation from the environment, triggering a temperature increase and a higher cost. We explore some solutions to this dilemma, showing how adiabatic computing is a very efficient strategy for fast operations, with a bounded cost only slightly above the fundamental limit.}

\correspondingauthor{\textsuperscript{1}ludovic.bellon@ens-lyon.fr}


\begin{abstract}
Landauer's principle makes a strong connection between information theory and thermodynamics by stating that erasing a one-bit memory at temperature $T_0$ requires an average energy larger than $\W_{LB}=k_BT_0 \ln2$, with $k_B$ Boltzmann's constant. This tiny limit has been saturated in model experiments using quasi-static processes. For faster operations, an overhead proportional to the processing speed and to the memory damping appears. In this article, we show that underdamped systems are a winning strategy to reduce this extra energetic cost. We prove both experimentally and theoretically that, in the limit of vanishing dissipation mechanisms in the memory, the physical system is thermally insulated from its environment during fast erasures, i.e. fast protocols are adiabatic as no heat is exchanged with the bath. Using a fast optimal erasure protocol we also show that these adiabatic processes produce a maximum adiabatic temperature $T_a=2T_0$, and that Landauer's bound for fast erasures in underdamped systems becomes the adiabatic bound: $\W_a = k_B T_0$.
\end{abstract}

\dates{This preprint was accepted for publication on July 25, 2023: PNAS {\bf 120} e2301742120}
\doi{\url{https://doi.org/10.1073/pnas.2301742120}}

\graphicspath{{Figures/}}

\begin{document}

\maketitle
\thispagestyle{firststyle}
\ifthenelse{\boolean{shortarticle}}{\ifthenelse{\boolean{singlecolumn}}{\abscontentformatted}{\abscontent}}{}

\dropcap{I}nformation is stored and processed in the material world, and as such is ruled by physics laws: handling information requires energy~\cite{Landauer-1961,Parrondo_sagawa,Orlov_book}. R. Landauer laid the foundations for the connection between information theory and thermodynamics by demonstrating theoretically the lower energy bound required to erase a one-bit memory: $\W_{LB}=k_BT_0 \ln2=\SI{3e-21}{J}$ at room temperature $T_0\sim\SI{300}{K}$, with $k_B$ Boltzmann's constant~\cite{Landauer-1961}. This energetic cost has an entropic origin: the number of possible states goes from two bit values (0 and 1) to only one value (the reset value, 0 for example). Since the entropy decreases at constant temperature $T_0$, the second law of thermodynamics implies that in average an energy has to be paid: this is Landauer's bound. This tiny limit has been experimentally illustrated, using quasi-static processes in model experiments~\cite{Berut2012, Berut2015, orl12, Bech2014, Gavrilov_EPL_2016, Finite_time_2020, Hong_nano_2016, mar16, Dago-2021}. When decreasing the duration of operations, an energy overhead proportional to the processing speed appears~\cite{Finite_time_2020, Berut2012,Aurell_2012,sek66,PhysRevE.92.032117}, and could explain why nowadays fast processors still consume orders of magnitude more energy than Landauer's bound. Indeed, on top of the average stochastic (or entropic) cost $\langle \W_S\rangle$, in finite time a second source of energy loss must be considered: the deterministic cost $\W_D$ of performing a fast erasure process in a viscous environment. Bypassing this overhead for fast operations would be a huge step towards efficient information processing. For this reason several new approaches have been explored, such as momentum computing~\cite{Crutchfield_2023} and Hamiltonian memories~\cite{Raz_2021}.

In this context underdamped systems~\cite{Dago-2021, Dago-2022-PRL, gieseler_levitated_2018, gieseler_non-equilibrium_2015} are a natural choice to minimize dissipation and to reduce in this way the deterministic cost $\W_D$ of fast erasures. However the reduction of the damping implies reducing the system coupling with the environment. Since this coupling rules the thermalization, the temperature of underdamped memories increases with the erasure speed. Thus the stochastic cost $\W_S$, linked to the system's temperature, rises ! The total energetic cost of erasing underdamped memories  therefore still presents an overhead to Landauer's bound, due both to the residual dissipation and to the warming effect~\cite{Dago-2022-PRL}. The key question is to measure, understand and minimize this warming effect along the overall energy cost of the fast erasure of these adiabatic memories.

\begin{figure*}[ht]
	\includegraphics[width=8.5cm]{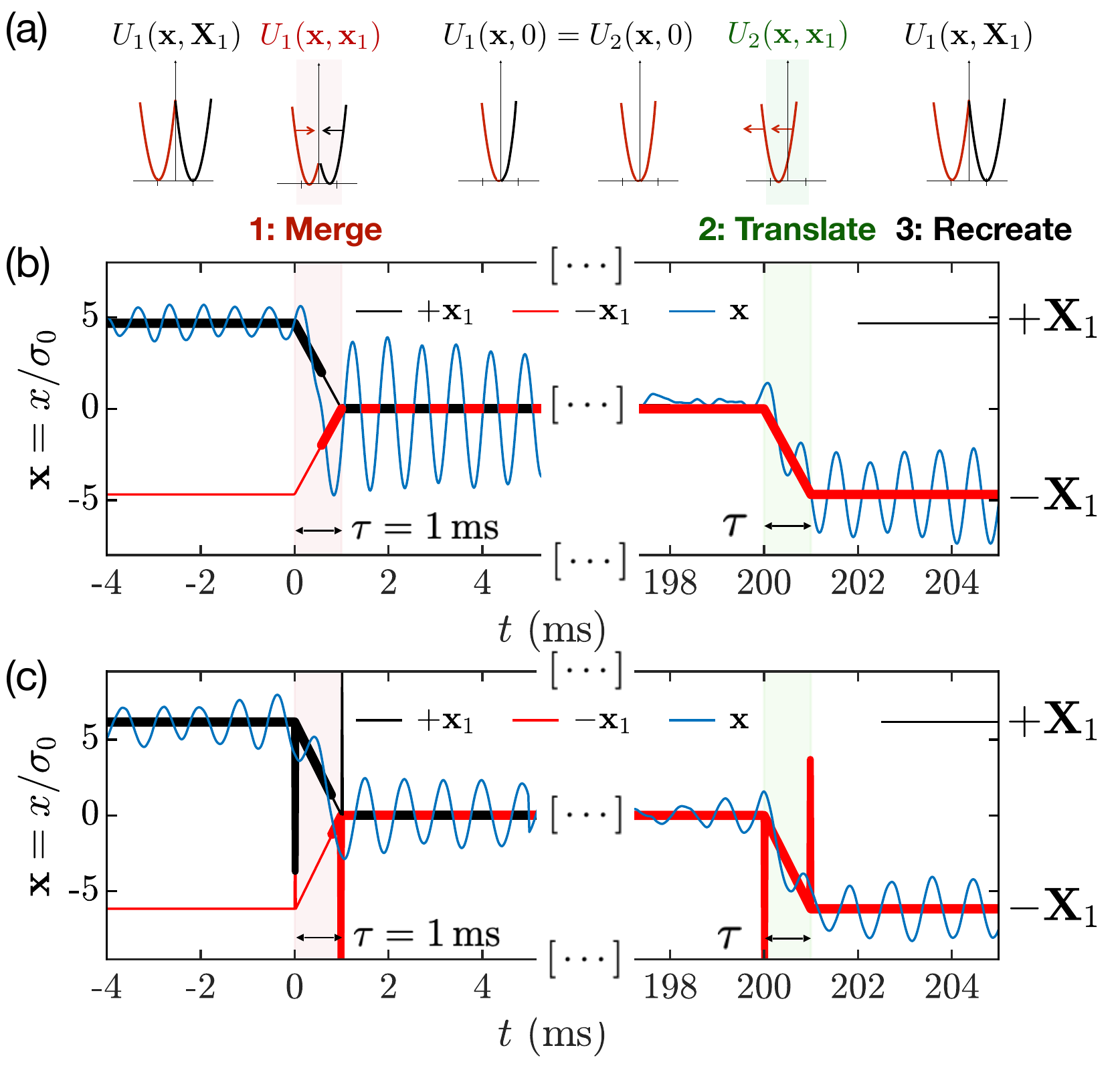} \hfill \includegraphics{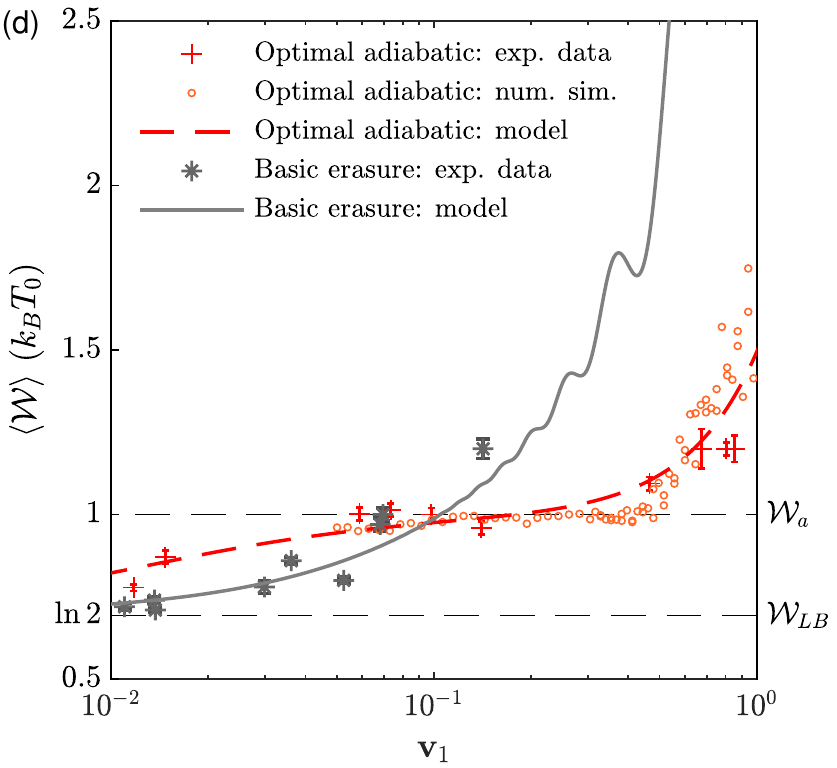} 
	\caption{\textbf{Erasure protocol and minimization of fast erasures cost using optimal adiabatic computing.} \textbf{(a) Schematic view.} Snapshots of the oscillator potential energy $U(\x,\x_1)$ during the erasure protocol, $\x$ being the oscillator position and $\pm\x_1$ the position of the center of the well(s), both expressed in units of $\sigma_0$, i.e. the equilibrium standard deviation of the position. We start in a double well bi-quadratic potential $U_1(\x,\x_1=\X_1)$, before executing the following steps: step 1 [Merge] consists in merging the two wells together into one well centered in $\x_1=0$, step 2 [Translate] consists in translating the single well $U_2(\x,\x_1)$ to the position $-\X_1$ of state 0, and finally step 3 [Recreate] consists in recreating the second well in position $+\X_1$ to get the initial potential $U_1(\x,\x_1=\X_1)$ back. \textbf{(b) Basic erasure driving.} In the basic erasure protocol, the wells centers are translated at constant speed $\mathbf v_1=\X_1/(\tau \omega_0)$ where $\tau$ is the erasure characteristic time, $\omega_0$ the angular frequency of the underdamped system in a single well, and $\X_1\sim 5$. One experimental trajectory in response to this basic driving is plotted in blue, for $\tau=\SI{1}{ms}$, corresponding to $\mathbf v_1=0.8$. The effective well center for this specific trajectory is highlighted by the thicker line for $\pm \x_1$.
	\textbf{(c) Optimal erasure driving.} All the ramps of $\x_1$ are replaced by optimal drivings~\cite{Gomez} as the one shown in Fig.~\ref{meanz}, materialized by the peaks in $\pm \x_1$ at the beginning and the end of each ramp. One experimental trajectory in response to this optimal driving is plotted in blue, again for $\mathbf v_1=0.8$: it has smaller fluctuations after the wells displacement than non optimal trajectories. 
	 \textbf{(d) Erasure energetic cost} The average work required to erase a 1-bit information is displayed versus the protocol speed $\mathbf v_1$. Optimal adiabatic erasure protocols are performed using a highly underdamped memory ($Q=80$). Both experimental (red crosses) and numerical simulation (red circles) results match the model (dashed red) and confirm the efficiency of optimal adiabatic computing to contain the energetic cost of fast erasures. For comparison, the energetic consumption previously obtained with non optimal erasure at low damping ($Q=10$)~\cite{Dago-2021,Dago-2022-PRL}, plotted in gray line (model) and stars (measurement), is much higher for fast erasures. Error bars on experimental data correspond to the statistical uncertainty (standard deviation divided by $\sqrt{N-1}$, with $N\gtrsim1000$ the number of analyzed trajectories).}
	\label{intro}
\end{figure*}

The purpose of this article is to answer this question by showing both experimentally and theoretically that in the limit of vanishing dissipation mechanisms in the memory, the physical system is thermally insulated from its environment, leading to the adiabatic erasure limit, i.e. with no heat exchanges with the bath. We prove that in this limit, the memory temperature rise saturates to a factor 2:
\begin{equation} \label{eq:Ta}
T_a=2T_0,
\end{equation}
corresponding to an adiabatic extension of Landauer's bound. The average work to erase 1-bit information is in this case:
\begin{equation} \label{eq:Wa}
\W_a = k_B T_0.
\end{equation}
Those two results are very general as long as the memory initial states 0 and 1 are equivalent and can be approximated by harmonic energy confining potentials. We demonstrate this limit by studying a highly underdamped memory made by a mechanical oscillator whose position is confined in a double well bi-quadratic potential. Because of the very weak coupling with the bath this memory can be considered adiabatic in the fast erasures regime. The protocol that we use is sketched in Fig.~\ref{intro}(a-c) and described in the captions. The energy cost of the erasure can be reduced by optimizing (i) the protocol time dependency (to lower the deterministic cost $\W_D$ linked to the residual dissipation at high speed) and (ii) the coupling of the memory to the bath (to contain the average stochastic cost $\langle \W_S \rangle$ at low coupling). The main results of this article are summarized in Fig.~\ref{intro}(d), where the energy costs of two erasure protocols (a standard and an optimal) are plotted versus the dimensionless erasure speed $\mathbf{v}_1$. The gray curves are obtained by applying the standard protocol used in Ref.~\cite{Dago-2021} on a low quality factor oscillator ($Q\sim10$). We clearly see that the dissipation rapidly increases for fast erasures. The red curves represent the results on a high quality factor resonator ($Q\sim80$) with an optimal protocol which minimizes the energy cost during wells displacements~\cite{Gomez}. This latter approach to erasure consumes at most twice $\W_{LB}$ even when the erasure speed approaches the thermal noise rms velocity of the system. Our results hint at low damping and inertia as promising ingredients in building fast and energy efficient information processing devices. 

In our experiment, the physical observable is the position $x$ of an underdamped micro-mechanical oscillator~\cite{Dago-2021,Dago-2022-PRL,Dago-2022-JStat} (in the form of a cantilever) characterized by its angular resonance frequency $\omega_0$, mass $m$, stiffness $k=m\omega_0^2$, and quality factor $Q$. The quality factor can be tuned by removing the air in the cantilever chamber, from $Q\sim10$ at atmospheric pressure to $Q\sim100$ in light vacuum (\SI{1}{mbar}). The position standard deviation $\sigma_0=\sqrt{k_B T_0/k}$ at equilibrium is used as the unit length: normalized position are written in bold font, eg. $\x=x/\sigma_0$. Similarly, the velocity variance $\omega_0\sigma_0=\sqrt{k_B T_0/m}$ at equilibrium is used to normalize the speed quantities, eg. $\mathbf{v}=\dot{x}/(\omega_0\sigma_0)$, and $k_BT_0$ is used to normalized energetic quantities, eg. $\bW=\W/(k_BT_0)$. The 1-bit information is encoded in the position $x$ using a double-well bi-quadratic potential, $U_1(x,x_1)= \frac{1}{2}k(| x | -x_1)^2$, with $x_1$ the user-controlled parameter, corresponding to the half-distance between wells and tuning the barrier height~\cite{Dago-2021,Dago-2022-JStat}. The 1-bit information is therefore state 0 or state 1 if the system is respectively confined in the left or right hand well of $U_1$. At rest, we use $x_1=X_1 \gtrsim 5 \sigma_0$, high enough to secure the initial 1-bit information.

Our erasure protocol is similar to the approach used in previous stochastic thermodynamics realizations~\cite{Berut2012, Berut2015, orl12, Bech2014, Gavrilov_EPL_2016, Finite_time_2020, Hong_nano_2016, mar16}: lower the barrier, tilt the potential towards the reset state, raise the barrier. We rename those 3 steps as \emph{Merge}, \emph{Translate}, and \emph{Recreate} (an empty well), illustrated in Fig.~\ref{intro}(a-c). Step 1 [Merge] consists in decreasing $x_1$ from $X_1$ to $0$ in a time $\tau$, thus translating the center of the wells at mean speed $\mathbf v_1=\X_1/(\omega_0\tau)$ till complete merging into a simple quadratic well. Step 2 [Translate] consists in translating the single well to state 0 position, using the potential $U_2(x,x_1)= \frac{1}{2}k(x + x_1)^2$, with $x_1$ increased from $0$ to $X_1$ in the same time $\tau$. Step 3 [Recreate] finally restores the potential $U_1(x,X_1)$, thus recreating an empty well on the right hand side. Optional rest times can be applied between steps [around $\SI{200}{ms}$ between steps 1 and 2 in Fig.~\ref{intro}(b,c)]. This procedure has a 100\% success rate as the memory always ends in state 0 independently of the initial state. Experimental details are described in \methods~\ref{supp_exp}.

\begin{figure}[htb]
	\includegraphics{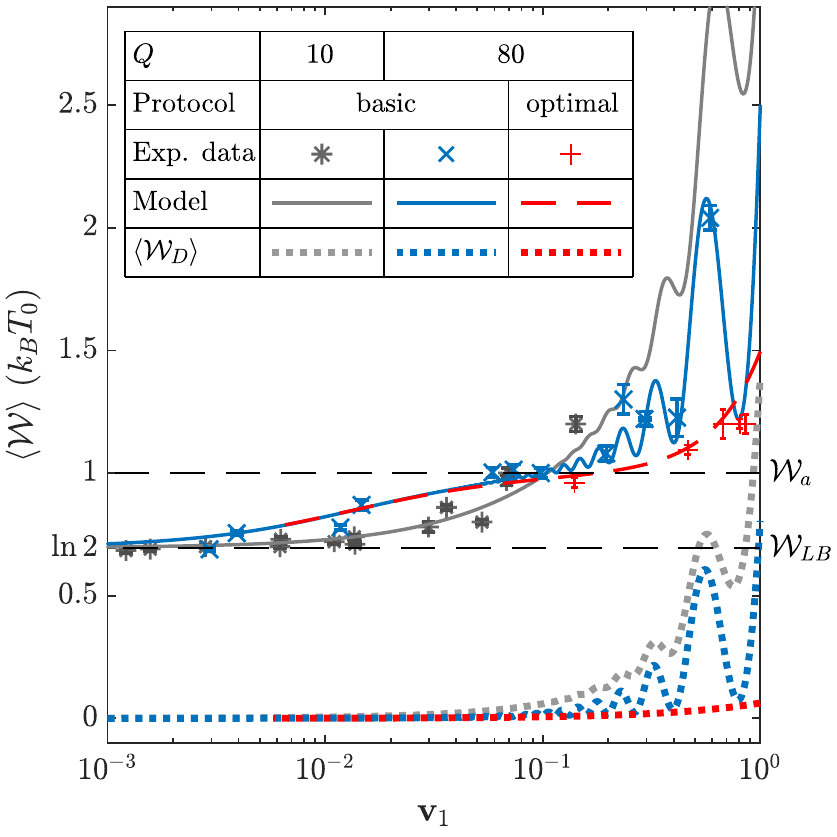}
	\caption{\textbf{Influence of the damping on the divergence from Landauer's bound for fast erasures}. As for the results at $Q=10$ (reproduced in grey from Ref.~\citenum{Dago-2021}), the experimental points (\textcolor{blue}{$\times$}) obtained at lower damping $Q=80$ (and still $\X_1\sim5$) are in good agreement with our model (blue line). We identify two speed regions in which the quality factor minimizing the erasure cost differs. In the moderate speeds region $\mathbf v_1<0.1$, $\W$ is minimized with $Q=10$, since at higher quality factor $Q=80$ what is won on the viscous work $\W_D$ is lost by the more substantial average warming cost $\langle\W_S\rangle$. On the contrary, for very fast erasures ($\mathbf v_1>0.1$), the overall cost is lower at $Q=80$, as it is ruled by the dissipative term only, the compression contribution being bounded by $\W_a=k_BT_0$. And indeed the translational motion cost from dissipation $\W_D$ is lower at very low damping ($Q=80$ in dotted blue line and $Q=10$ in dotted grey line). A further economy in the erasure cost can be attained by applying an optimal translational driving (red), which suppresses the oscillations at high speed: force kicks to set the system in motion or brake it are applied to kill transients oscillations during and after the translations. Error bars on experimental data correspond to the statistical uncertainty (standard deviation divided by $\sqrt{N-1}$, with $N\gtrsim1000$ the number of analyzed trajectories).}
	\label{Approach_vide_W2}
\end{figure}

In order to minimize the total erasure cost, each step should be optimized. The easiest one is step 3 [Recreate]: it has no energy cost, since it modifies the potential of a statistically unreachable part of the phase space. We then proceed with step 2 [Translate], which is deterministic: the system evolves in an harmonic well, it is therefore linear and sums the deterministic response to the driving and to the stochastic response to the thermal noise of the thermostat at temperature $T_0$ --- the driving does not affect the stochastic thermal properties of the system, and vice-versa. The average energetic cost $\W_{D}$ is thus purely deterministic and can be analytically computed~\cite{Dago-2022-PRL}. If we simply \emph{translate the well at a constant velocity} $\mathbf v_1$, as in Refs.~\citenum{Dago-2021,Dago-2022-PRL}, we obtain the dotted grey ($Q=10$) and dotted blue ($Q=80$) curves of Fig.~\ref{Approach_vide_W2}. Unsurprisingly, a lower dissipation (higher $Q$) results in a lower work. We notice that the deterministic work $\W_{D,\textrm{bas}}$ during this \emph{basic driving} oscillates when $\mathbf v_1$ increases. For fast protocols, the commensurability of $\tau$ with the oscillator period matters: an integer number of periods allows the system to end the translation at zero mean velocity, while any deviation leaves the system with a detrimental extra kinetic energy.

\begin{figure}[!h]
	\includegraphics[width=8.5cm]{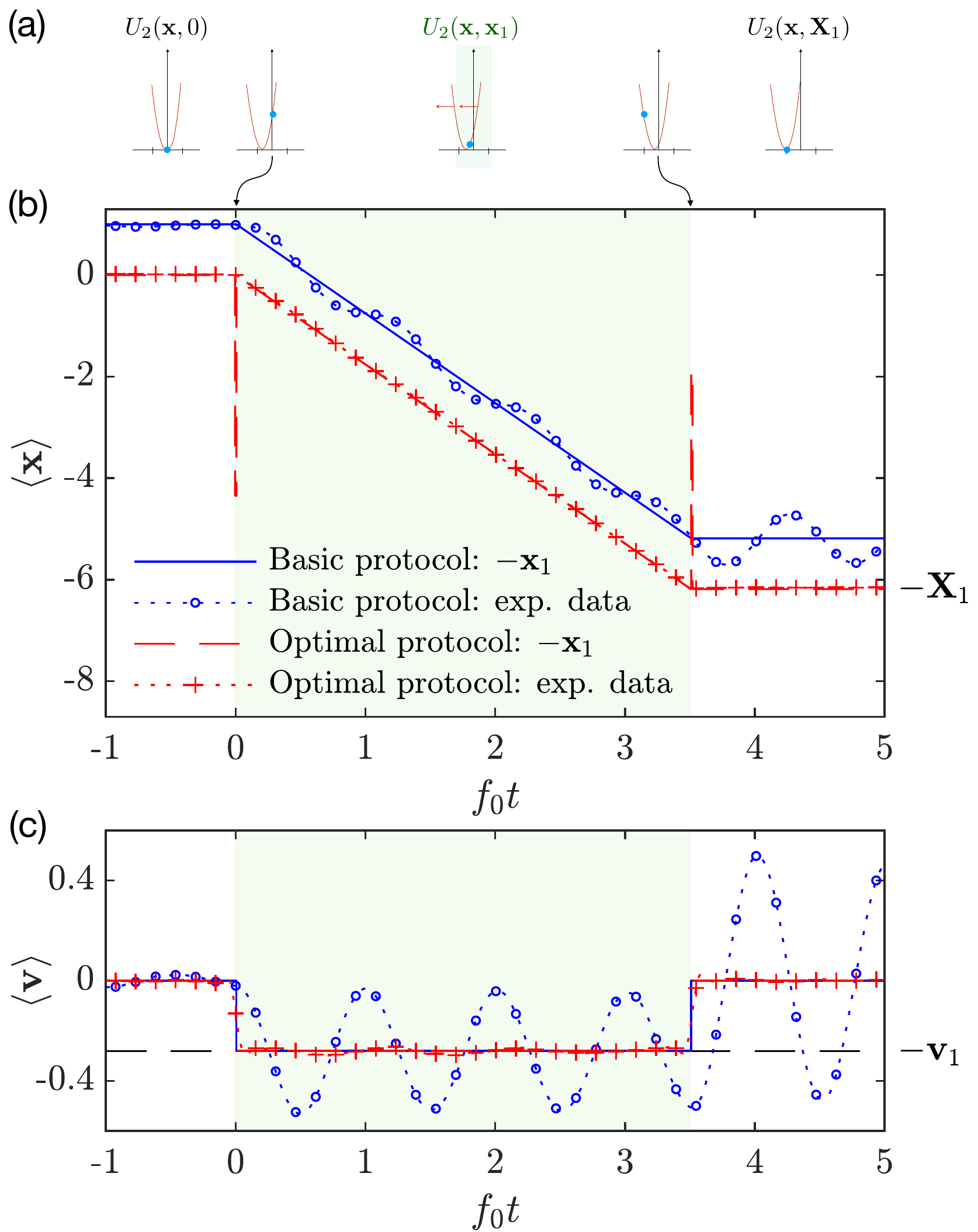}
	\caption{\textbf{Optimal versus basic translational driving for step 2.} The basic driving consists in simple ramps to change $\x_1$ from $0$ to $-\X_1=-6.2$ in $\tau=\SI{2.8}{ms}=3.5/f_0$ (corresponding to $\mathbf v_1=0.28$). The optimal driving is computed from Ref.~\citenum{Gomez} to move along the same ramp an underdamped harmonic oscillator of quality factor $Q=80$ at a minimal cost given in \eqref{W2opt}. \textbf{(a)~Schematic view:} snapshots of the potential $U(\x,\x_1)$ of step 2, as in Fig.~\ref{intro}(a). \textbf{(b) Driving $-\x_1$ and average response in position $\langle \x \rangle$.} The basic protocol is plotted in blue (and shifted vertically for clarity), the optimal one in red. The latter presents delta like peaks that force the mean velocity $\langle \mathbf{v} \rangle$ to jump at the beginning of the process from zero to the optimal constant speed, and back to 0 at the end of the translation [see panel (c)]. The center of the well is slightly ahead of the oscillator during the translation, to compensate for the small remaining drag force (this effect is invisible here with $Q=80$). This optimal translational driving triggers no transient oscillations, as shown by the average on 2000 trajectories (\textcolor{red}{$+$}), by opposition to the basic translation (\textcolor{blue}{$\circ$}). \textbf{(c) Average speed.} The optimal driving successfully cancels the final velocity on average. The response to the basic driving shows an important transient in the velocity degree of freedom after the ramp. The effect is magnified here by the high speed, and because $\tau$ is not a integer number of the oscillator period $1/f_0$.}
	\label{meanz}
\end{figure}

For a given set of protocol parameters ($X_1$, $\tau$), one can avoid those oscillations with an \emph{optimal driving} scheme. This topic is studied in Refs.~\citenum{Gomez, Dellago}, and explicit protocols are available~\cite{Gomez}. In short, the driving is now tuned so that the deterministic motion of \emph{the oscillator is at constant speed}, getting rid of any transient oscillations, as displayed in Fig.~\ref{meanz}. Initial acceleration and final deceleration are performed by applying force peaks (kicks) at the initial and final instants of the ramp, such as the ones displayed in Fig.~\ref{intro}(c) and \ref{meanz}. As demonstrated in Fig.~\ref{meanz}, this optimal driving successfully suppresses the oscillations triggered in both degrees of freedom (position and velocity) during and after the ramp when using a basic driving. It results in the lowest possible work required to move the underdamped harmonic oscillator of a distance $\X_1$ in a time $\tau$~\cite{Gomez}: 
\begin{align}
 \bW_{D,\textrm{opt}}=\frac{\X_1^2}{2+Q\tau\omega_0}= \frac{\X_1^2}{2+Q\X_1/\mathbf{v}_1}. \label{W2opt}
\end{align}
Let us point out that contrary to $\W_{D,\textrm{bas}}$ that oscillates when $\mathbf{v}_1$ increases, $\W_{D,\textrm{opt}}$ slowly and monotonically grows (dotted red curve on Fig.~\ref{Approach_vide_W2}). As a last demonstration of the optimal translational driving efficiency, Fig.~\ref{dW} in \methods\ compares the energy consumption during basic and optimal protocol. To summarize, we use optimal driving to get rid of transients oscillations during deterministic motions: it minimizes the energetic cost associated to viscous damping, especially at high velocity.

We finally focus on the energy exchanges during step 1 [Merge], which are partly deterministic and partly stochastic. Indeed at the beginning, when the two wells are moved towards the center, the system remains in its initial well without noticing and exploring the other well. Thus the behavior is equivalent to the translation of step 2: it is initially deterministic. Using large $Q$ and an optimal driving scheme, as for step 2, reduces this initial deterministic contribution to its minimum. However, when the system starts exploring the two wells, step 1 becomes stochastic, and is equivalent to a compression of the position space accessible to the oscillator. Similarly to the compression of a gas, this step modifies some stochastic properties of the system: the temperature rises. In this analogy, the average stochastic cost $\langle\W_S\rangle$ of this compression is then expected to increase with the temperature, and should be evaluated. A statistical physics model encompassing all those effects is developed in Ref.~\citenum{Dago-2022-PRL}. 

In order to minimize the erasure cost, one has to act on both the deterministic costs $\W_D$ during steps 1 and 2, and on the average stochastic cost $\langle\W_S\rangle$ of step 1, materialized through the warming of the memory. According to Eq.~\ref{W2opt}, optimal driving with low dissipation memories are desirable to minimize $\W_D$. We therefore perform experiments in vacuum to lower damping, and additionally use optimal driving for every translations of the well(s): this is our \emph{optimal erasure protocol}\footnote{Ref.~\citenum{Finite_time_2020} defines an optimal erasure designed for over-damped systems that cuts the dissipative cost by reducing the translational driving amplitude at the expense of an out-of-equilibrium final state. Though the final state after step 1 [Merge] is also out-of-equilibrium, optimal protocol in the present article stands for optimized ramps of the wells center.}, illustrated in red (dashed curves and experimental/numerical simulation data) in Figs.~\ref{intro}(d) and \ref{Approach_vide_W2}. The model adapted from Ref.~\citenum{Dago-2022-PRL} and detailed in \methods~\ref{supp_thermo} and \ref{supp_model} is in good agreement with all the experimental data. We confirm that optimal erasure at high $Q$ lowers the cost of fast erasures, at $\mathbf v_1>0.1$.

Nevertheless, moderate damping ($Q=10$) is still cheaper at lower speeds. This counter-intuitive result implies that $\W_D$ and $\langle\W_S\rangle$ have opposite behaviours when $Q$ grows. While $\W_D$ is reduced, so are the heat exchanges with the thermostat. Indeed, the dimensionless parameter $2\pi/Q$ of the oscillator represents the amount of energy exchanged with the thermal bath during one oscillation period, normalized by the total energy of the oscillator. At low damping ($Q \gg 1$), energy exchanges are thus very small: the memory is thermally insulated on short time scales. The low dissipation limit then corresponds to the adiabatic limit for the oscillator: the work influx cannot be immediately balanced by heat dissipation, and the warming of the memory is large. Since $\langle\W_S\rangle$ sums an entropic part, $\W_{LB}=k_BT_0\ln{2}$, and an extra-cost originated in the warming of the memory, the adiabatic limit results in an increase of the average stochastic work of erasure $\langle\W_S\rangle$. At low speed ($\mathbf v_1<0.1$), $\W_D$ is small and $\langle\W_S\rangle$ prevails: moderate damping results in cheaper erasures. In \methods~\ref{supp_model}, we provide the average erasure cost map as a function of $Q$ and $\mathbf{v}_1$, and show that its minimum is reached for $Q\sim5$. For fast erasures ($\mathbf v_1>0.1$), the increase of $\W_D$ dominates and the reduction of dissipation is a winning strategy: this is the adiabatic computing regime. This suggests that $\langle\W_S\rangle$ is bounded in this regime, as anticipated by Eq.~\ref{eq:Wa} in the introduction: $\W_{LB} \leq \langle\W_S\rangle \leq \W_a$. We experimentally and theoretically study this key point in the following paragraphs. 

In the limit of high quality factors, heat exchanges are slow, so that the average stochastic work $\langle\W_S\rangle$ due to a fast compression in step 1 [Merge] results in the rise of the temperature $T$ of the memory. $T$ is defined as the kinetic temperature, using the (non-deterministic) average kinetic energy (see \methods~\ref{supp_temp} for details):
\begin{align}
\langle K\rangle-K_D=\frac{1}{2}m \langle v^2\rangle - \frac{1}{2}m\langle v \rangle ^2=\frac{1}{2}m \sigma_v^2=\frac{1}{2} k_B T
\end{align}
Let us emphasize that the notion of kinetic temperature holds for a Gaussian Probability Distribution Function (PDF) of speed (such as the Boltzmann distribution at equilibrium), when the speed variance and mean are the only moments of the distribution. We restrict to $\mathbf v_1 < 0.5$ to tackle the memory's temperature evolution: as we address in \methods~\ref{PDF}, for larger $\mathbf v_1$, the system has not the time to explore the whole phase space and a Gaussian distribution no longer properly approximate the speed PDF.

\begin{figure}[htb]
	\includegraphics{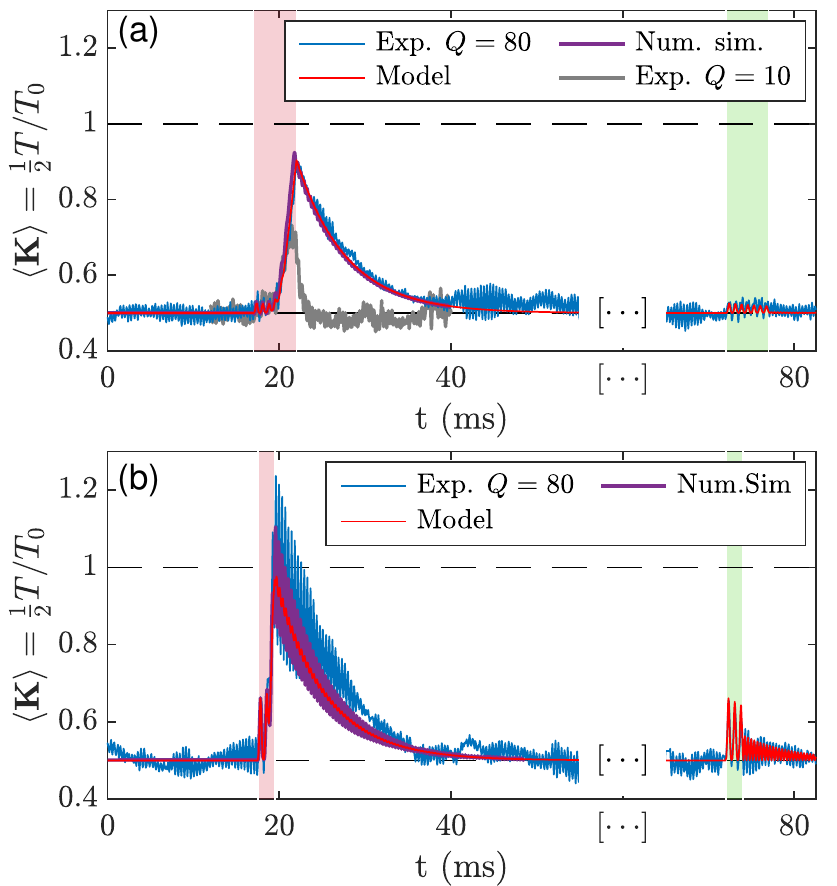}
	\caption{\textbf{(a) Kinetic energy during a fast erasure process ($\mathbf v_1=0.12$) at high quality factor ($Q=80$)}. Step 1 in red background lasts $\tau=\SI{5}{ms}$ (with $\X_1=5$) and results in a strong temperature rise visible on the kinetic energy profile: $\langle K \rangle$ culminates at $\SI{0.92}{k_BT_0}$, ie at $92\%$ of the adiabatic limit $K_a=k_BT_0$. At the end of step 1 [Merge], the system thermalizes with the surrounding bath in $\tau_\mathrm{relax}\sim\SI{20}{ms}$ so that the kinetic energy relaxes to its equilibrium value $K_{\textrm{eq}} = \frac{1}{2} k_B T_0$. Then, the translational motion of duration $\tau$ (step 2 in green background) only produces tiny oscillations. The experimental curve averaged from $N\gtrsim1000$ trajectories (blue), nicely matches the model without any tunable parameters (red) and the simulation result for step 1 (purple) obtained from $N_{\textrm{simu}}=10^5$ simulated trajectories. The measurement at the same speed but for $Q=10$ is reported in grey for comparison. \textbf{(b) Kinetic energy in the adiabatic limit: $\mathbf v_1=0.3$ and $Q=80$}. This time the adiabatic limit is reached during the compression of duration $\tau=\SI{2}{ms}$ as predicted by the model and confirmed by the numerical simulation. The average kinetic energy does not exceed $\K_a=k_BT_0$ except for the transient deterministic contribution $K_D$ during this basic protocol. Step 2 basic translational motion also triggers larger oscillations than those observed in (a).}
	\label{Kadiab}
\end{figure} 

To investigate the temperature evolution during an erasure, we thus look at the speed variance evolution from the thousands recorded trajectories for different erasure speeds and quality factor. Actually for the erasure speeds under study the deterministic contribution can be assumed negligible ($K_D\sim \frac{1}{2} m v_1^2 \ll \langle K\rangle$) so that the kinetic temperature evolution is directly visible through the kinetic energy: $\langle \bK \rangle=\frac{1}{2} T/T_0$. In Fig.~\ref{Kadiab}(a) we compare the time trace of $\langle \bK\rangle$ in air and vacuum during basic erasures at the same speed $\mathbf v_1=0.12$ ($\tau=\SI{5}{ms}$). It demonstrates that the larger the $Q$, the larger the temperature rise: $\langle K\rangle$ peaks at $\SI{0.92}{k_BT_0}$ for $Q=80$, compared to $\SI{0.72}{k_BT_0}$ for $Q=10$. This peak is followed by an exponential relaxation to equilibrium, which is faster for a higher dissipation (thus for $Q=10$ here). If we more than double the erasure speed at $ \mathbf v_1=0.3$ ($\tau=\SI{2}{ms}$), we observe in Fig.~\ref{Kadiab}(b) that the warming intensity is further increased, but does not exceed $\K_a=k_BT_0$ (except for transient oscillations at $2\omega_0$).

\begin{figure}[htb]
	\includegraphics{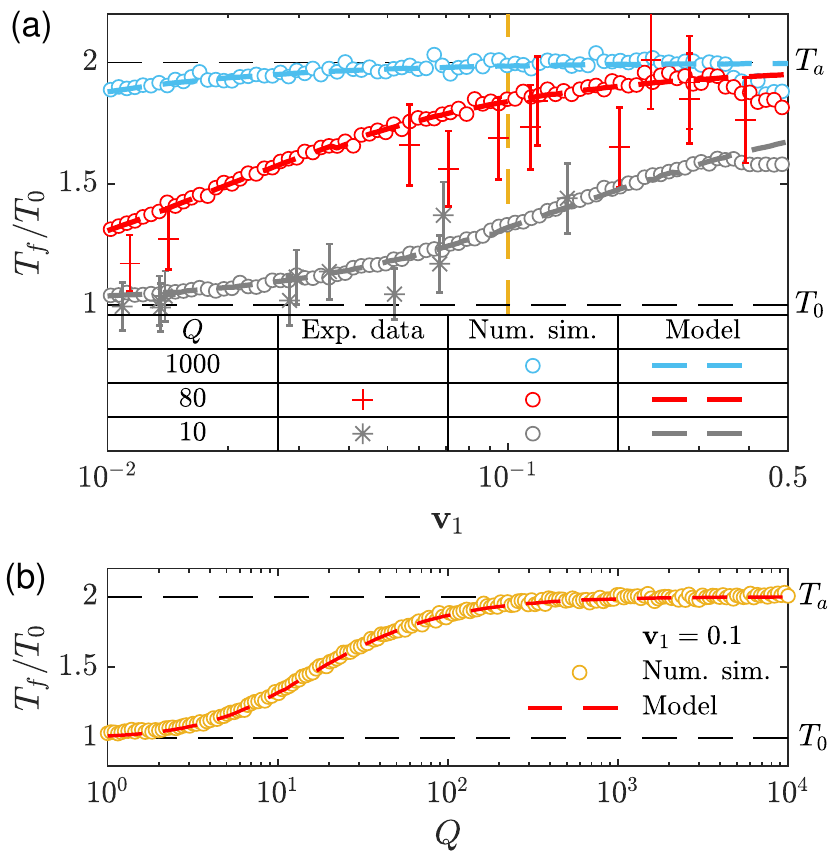}
	\caption{\textbf{(a) Temperature rise amplitude} $T_f/T_0$ 	as a function of the optimal erasure speed $\mathbf v_1=\X_1/(\tau \omega_0)$ for different quality factor, $Q=10$ (gray), $Q=80$ (red) and $Q=1000$ (blue). Experimental results for $Q\sim 10$ (\textcolor{gray}{\ding{83}}) and $Q\sim 80$ (\textcolor{red}{$+$}) are obtained by averaging the speed variance $\sigma_v^2$ computed from\ $N\gtrsim1000$ trajectories during half the oscillator period at the end of step 1 (corresponding to one transient oscillation of squared quantities such as $\sigma_v^2$). The error bars correspond to the statistical uncertainty (standard deviation divided by $\sqrt{N-1}$). Simulation results ($\circ$) are deduced similarly from $10^4$ numerically simulated trajectories. Experiments and numerical simulations are in very good agreement with the model (dashed lines) with no adjustable parameters. \textbf{(b) Approach to the adiabatic limit} of the temperature rise amplitude $T_f$, when the quality factor $Q$ increases for fast erasures ($\mathbf v_1=0.1$), corresponding to a cut of panel (a) along the vertical orange dashed line. Simulation results (\textcolor{orange}{$\circ$}) obtained again by averaging $\sigma_v^2$ at the end of step 1 perfectly match the model (dashed line). Both tend to the adiabatic limit $T_a=2T_0$ when the dissipation is removed.} \label{Kmax}
\end{figure}

All the above experimental results are supported by numerical simulations. $\K_a$ corresponds to the adiabatic limit on the temperature, $T_a=2T_0$, which is reached for large quality factor when the erasure is fast enough. The approach to the adiabatic limit is illustrated on Fig.~\ref{Kmax}: the kinetic temperature reached at the end of step 1 for optimal erasures, $T_f=T(t=\tau)$, is plotted as a function of the erasure speed for different $Q$ on Fig.~\ref{Kmax}(a), and as a function of $Q$ for $\mathbf v_1=0.1$ on Fig.~\ref{Kmax}(b). Experimental and simulation results match and show that $T_f$ approaches the adiabatic limit $T_a=2T_0$ at erasure speeds that decrease at high quality factor. Indeed in those limits, the heat exchanges are negligible during the compression of step 1 (hence becoming an adiabatic compression), and delayed to the relaxation period after step 1. Let us point out again that when $\mathbf v_1 \gtrsim 0.5$, the concept of kinetic temperature is no longer well defined, and the simulation data based on the computation of the speed variance progressively become meaningless.
However, if we stick to fast erasures allowing to define the memory's temperature (setting $\mathbf v_1 =0.1$ for example), we see on Fig.~\ref{Kmax}(b) that $T_f \to T_a$ when $Q \to \infty$. We model those behaviors in further details in the next paragraph.

In Ref.~\citenum{Dago-2022-PRL}, we propose an efficient theoretical framework to predict the energy exchanges and explore the fast information erasure cost. The model only requires the system parameters ($\omega_0$ and $Q$) and the protocol ones ($\X_1$ and $\tau$) to estimate the average erasure cost: the latter is then computed as a function of $Q$ and $\mathbf v_1=\X_1/(\omega_0 \tau$). The model relies on the gaussian ansatz for the speed and position PDF (see \methods~\ section~\ref{PDF}) accurate for $\mathbf v_1 \lesssim 0.5$ and valid in first approximation until $\mathbf v_1 =1$. Fig.~\ref{Kadiab} demonstrates the accuracy of this approach (red line) to predict the kinetic energy profile (mirror of the temperature evolution) in experimental and simulation results. In addition, Fig.~\ref{Kmax} also validate the model prediction of the temperature rise amplitude as a function of $\mathbf v_1$ and $Q$, by successfully comparing it to simulation and experimental results.

However, the adiabatic limit case can be described easily without going through the whole model complex solving: we propose in \methods~\ref{sec_phase_volume} an alternative demonstration to the one of Ref.~\citenum{Dago-2022-PRL}. In a nutshell, this proof is based on Liouville's theorem~\cite{Peliti_book,Tong_book,Neishtadt_2019}, which states that the phase space volume is conserved during a transformation of an Hamiltonian system. In the adiabatic limit ($Q \to \infty$), step 1 [Merge] occurs in a time too short to exchange energy with the bath: $\tau\ll\tau_\mathrm{relax}=2 Q/\omega_0$, {\it i.e.} $\mathbf{v}_1 \gg \X_1/(2Q)$. The system evolution is then Hamiltonian: stochasticity comes from initial conditions only. In the phase space ($x$, $v$), the surface corresponding to energies smaller than an initial energy $E_i$ is enclosed by two ellipses, of total area $4\pi E_i /\sqrt{mk}$, as illustrated in Fig.~\ref{Liouville}. This area is conserved, and as long as the transformation is not too fast ($\mathbf{v}_1\ll 1$), the surface corresponding to the final energy $E_f$ is enclosed by a single ellipse of area $2\pi E_f / \sqrt{mk}$. For protocol speeds $\mathbf{v}_1$ such that $\X_1/(2Q) \ll \mathbf{v}_1 \ll 1$, Liouville's theorem then directly imply that $E_f=2E_i$: the energy doubles due to compression. Since the system starts and ends in a quadratic potential, the average energy in the initial and final states satisfies $\langle E_i \rangle /T_0=\langle E_f \rangle /T_f=k_B$, hence $T_f=T_a=2T_0$, as anticipated in Eq.~\ref{eq:Ta}. The average work performed is also readily computed as $\W_a = \langle E_f \rangle - \langle E_i \rangle = \langle E_i \rangle = k_B T_0$, as anticipated in Eq.~\ref{eq:Wa}. To summarize, an adiabatic compression results in doubling the system temperature (and therefore the kinetic energy, $K_a=k_BT_0$) and requires on average $\W_a=k_BT_0$ of work, as the conservation of the phase space volume enslaves the variations of the energy and of the temperature to those of the volume (see Ref.~\citenum{Raz_2021} for a discussion on Hamiltonian memories). Let us point out that the factor 2 in the phase space volume is a direct consequence of the 1 bit encoding and of the erasure initial and final states. In the general case, the two wells of equivalent shape are initially equally populated, and the erasure displaces half this population in the target state. Only the expression of the energy enclosed in this phase space volume and its relation to the temperature can be modified outside the harmonic approximation of the energy in state 0 and 1. Nevertheless the same reasoning can be easily adapted to any non harmonic potential wells, as shown in the ancillary files.

As summarized in Fig.~\ref{intro}, this article gathers experimental, numerical simulation and theoretical results demonstrating that an optimal adiabatic protocol is the cheapest way to erase 1 bit of information at high speed. Such protocol relies on two complementary approaches. The first consists in reducing the damping to reach the adiabatic limit of the warming cost (stochastic part). The second implements an optimal protocol minimizing the remaining dissipative cost (deterministic part). The combination of these strategies lowers substantially the cost of information processing for $\mathbf v_1>0.1$. The study of the adiabatic limit proves experimentally and theoretically that the warming of the memory is bounded by $T_a=2T_0$ (for erasure speeds allowing to define the kinetic temperature). This temperature rise is due to the delayed heat flow towards the thermostat, which occurs only slowly in the limit of low damping.

Since the system has to relax to equilibrium after the erasure, this strategy takes advantage of a non equilibrium final state to limit the energetic cost, similarly to other approaches~\cite{Finite_time_2020,Raz_2021,Gomez}.
Nevertheless, in order to apply optimal adiabatic computing concretely and in particular allow fast successive erasures without unwanted and uncontrolled temperature rise~\cite{Dago-thesis}, one would need to find a way to shortcut this thermalization. There are several possibilities to reach this goal. A first one is the use of shortcut to adiabaticity (STA) techniques which allow the system to bypass the long relaxation time and to relax very fast. STA techniques consist in using an optimal protocol which minimizes the energy needed for speeding up the relaxation (see for example Refs.~\citenum{ChupeauUD,PlataTrizac2020}). A second possibility is to increase the damping during the cooling step only (for example artificially using a velocity feedback, or by a quickly tunable dissipation mechanism), reducing in this way the relaxation time. A third possibility is to increase the heat capacity of the memory so that its temperature raises less during the work influx. Finally the warming up of the memory could be turned into a feature: the thermal energy in the memory is not lost yet in the thermostat at the end of step 1. One could imagine a specific protocol that takes advantage of the two temperatures available: the thermostat at $T_0$ and the hot memory at around $2T_0$. If such a thermal engine resulting in work extraction could be implemented, then cooling the memory would lower the actual overall cost of the erasure, potentially allowing one to meet Landauer's bound even for fast erasures.

%

\matmethods{
\newcounter{AppendixEquation}
\newcounter{AppendixFigure}
%
%
\stepcounter{AppendixEquation}
\stepcounter{AppendixFigure}
\renewcommand{\thefigure}{M\arabic{figure}}
\renewcommand{\theequation}{M\arabic{equation}}
\section{Data availability}
The data that support the findings of this study are openly available in Zenodo~\cite{Dago-2023-DatasetPNAS}.

\section{Experimental set up: virtual potential created by a digital feedback}
\label{supp_exp}

The experimental setup is sketched on Fig.~\ref{setup}(a). It is very similar to the one described in Refs.~\citenum{Dago-2021,Dago-2022-PRL,Dago-2022-JStat}: the first oscillation mode of a conductive cantilever is used as an harmonic oscillator, whose equilibrium position is tuned using an external voltage $V_0+V$ applied between the lever and a facing electrode. Its position $x$ is measured with great accuracy with a differential interferometer~\cite{Bellon-2002,Paolino2013}. To create the double well, we use a fast feedback loop to enslave the voltage to the sign of $x$: $V=\pm V_1$. Using $V_1 \ll V_0$ linearizes the electrostatic force effect and creates two symmetric harmonic wells centered in $\pm x_1$, where $x_1\propto V_1$. With respect to our previous experiments~\cite{Dago-2022-JStat}, there are two differences which we overview in the next two paragraphs: measurements in vacuum, and a digital feedback loop.

\begin{figure}[!b]
	\includegraphics[width=\columnwidth]{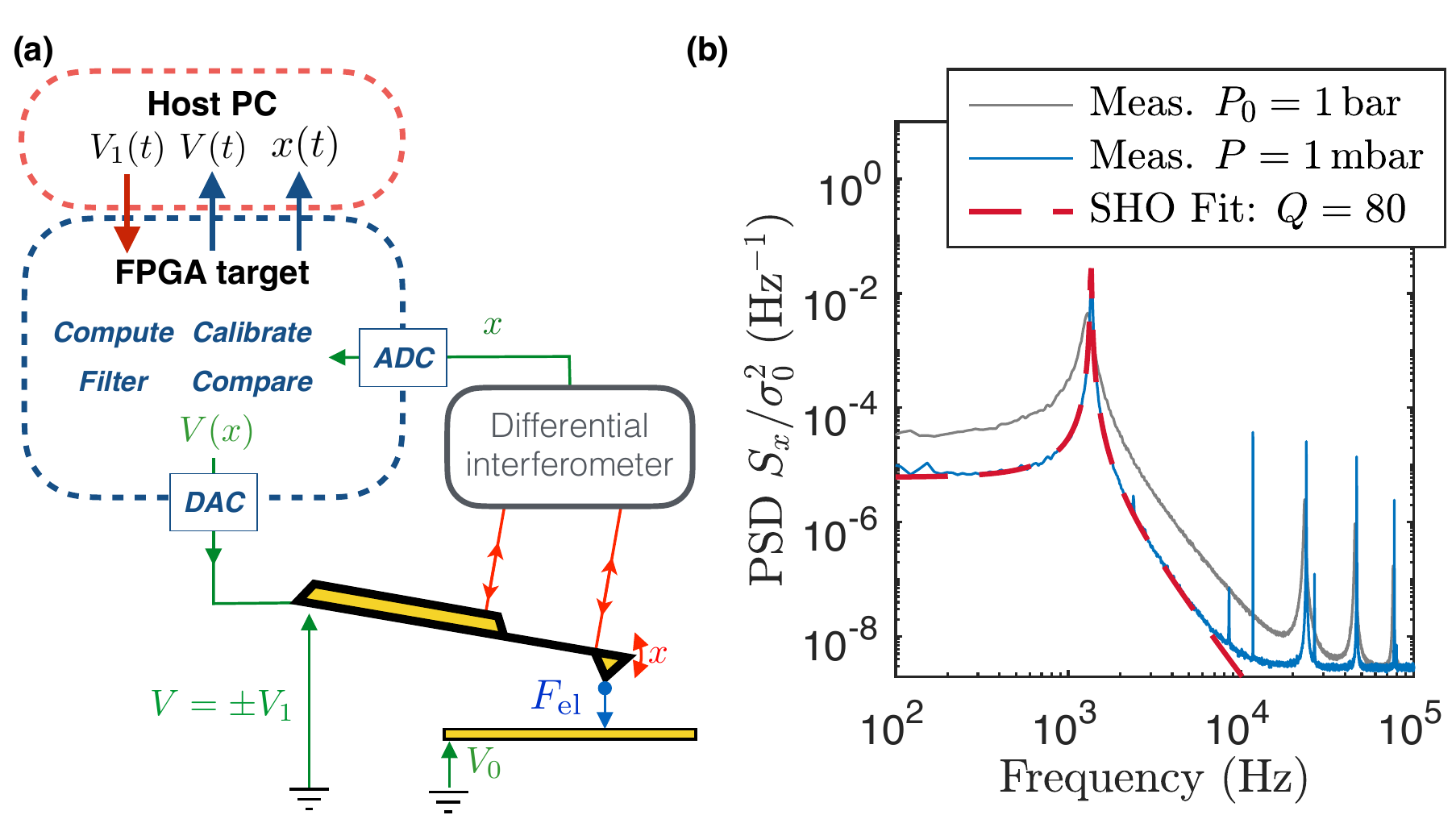}
	\caption{ \textbf{(a) Schematic diagram of the experiment}: the resonator is a conductive cantilever sketched in yellow. Signals from the differential interferometer are digitized and processed in real time using a FPGA clocked at $\SI{100}{MHz}$. The resulting calibrated position $x(t)$, filtered to remove high frequency noise, is compared to 0 (except during step 2) to compute the output voltage $V=\pm V_1(t)$ and create via the electrostatic force $F_\mathrm{el}$ a virtual double well potential. The feedback delay is around $\SI{1}{\mu s}$. All the signals are saved in the host PC. \textbf{(b) Thermal noise spectrum at very low damping}: the PSD of the cantilever deflection, with no feedback, is plotted as a function of frequency. The laser beam is focused on the node of the second deflexion mode to suppress its contribution. We compare the PSD in vacuum ($P=\SI{1}{mbar}$, blue) with the one at ambiant pressure ($P_0=\SI{1}{bar}$, grey): removing the viscous dissipation increases the amplitude and the sharpness of the resonances. The best fit by the theoretical spectrum of a Simple Harmonic Oscillator (SHO, dashed red line) confirms that up to $\SI{10}{kHz}$ the cantilever behaves in vacuum like a SHO at $f_0=\SI{1353}{Hz}$, with a quality factor $Q=80$.} \label{setup}
\end{figure}

We perform some measurements in light vacuum to increase the quality factor of the resonance from $Q\sim10$ in air to $Q\sim100$ at $\SI{1}{mbar}$. As displayed in Fig.~\ref{setup}(b), the measured Power Spectrum Density (PSD) of the deflection $x$ of the cantilever in vacuum in a single well is very well fitted by the thermal noise spectrum of a Simple Harmonic Oscillator (SHO), up to frequencies 10 times larger than its resonance frequency $f_0$. From the fit, we extract $f_0=\SI{1.3}{kHz}$, $Q=80$, $\sigma_0 = \sqrt{k_BT_0/k} = \SI{0.8}{nm}$. Several cantilevers have been used during our experiments, with slightly different values for $f_0$ and $\sigma_0$ (variations within $\sim20\%$ of those values). We therefore use normalized values most of the time to compare results between different experimental runs.

For such high quality factor, the analogical implementation of the feedback loop used in Refs.~\citenum{Dago-2021} and~\citenum{Dago-2022-PRL} does not meet anymore the experimental requirements to create a clean virtual potential~\cite{Dago-2022-JStat}. Indeed, the feedback response time has to be much lower when the system is not slowed down by the surrounding air anymore. We therefore turn to a digital feedback loop, based on a FPGA (Field Programmable Gate Array), achieving a response time of $\SI{1}{\mu s}$. This improved set-up better suited for high quality factors allows us to explore a wider panel of erasure speeds, up to $\tau\sim\SI{1}{ms}$. Besides, all the calibration, feedback and acquisition steps are implemented in the FPGA: after the calibration, the measured position $x$ is filtered and compared to $0$ to compute the output voltage $V(x,t) = V_1(t) x/|x| $, where $V_1(t)$ is shaped according to the chosen protocol. All the data are saved in parallel on the host PC. The setup is also programmed to compensate for slow drift of the zero of the interferometer, allowing more reliable lengthy measurements.
 
\section{Underdamped stochastic thermodynamics}
\label{supp_thermo}
We consider a Brownian system of mass $m$ in a bath at temperature $T_0$ characterized by its position $x$ and velocity $v=\dot{x}$. Its dynamic into a potential energy $U(x)$ is described by the 1-dimension Langevin equation, 
\begin{equation}\label{Langevin1D}
m \ddot{x} + \gamma \dot{x}=- \frac{\partial U}{\partial x}+ \sqrt{2 k_B T_0 \gamma}\xi(t).
\end{equation}
with $\gamma$ the damping coefficient of the environment ($\gamma=m\omega_0/Q$ in our case), and $\xi(t)$ a $\delta$-correlated Gaussian white noise:
\begin{equation} \label{xi}
\langle\xi(t)\xi(t+t')\rangle=\delta(t')
\end{equation}
modeling the thermal noise stochastic driving.
 
We introduce the kinetic energy $\K=\frac{1}{2}mv^2$. The equipartition gives the kinetic energy average value at equilibrium (as the potential does not depend on $v$):
 \begin{align}
\langle \K \rangle &=\frac{1}{2}k_BT_0.
\label{Eqpart_Ec}
\end{align}
As the total energy is worth $E=U+\K$, the energy balance equation writes: 
\begin{equation} \label{EqBalance}
\frac{dK}{dt}+\frac{dU}{dt}=\frac{d\W}{dt}-\frac{d\Q}{dt},
\end{equation} 
with $\W$ and $\Q$ the stochastic work and heat defined by~\cite{sek10,sek66,Seifert_2012,Jarzynski_2011,Ciliberto_PRX,Dago-2022-PRL}:
\begin{align}
\frac{d\W}{dt}&\equiv \frac{\partial U}{\partial x_1} \dot x_1 \label{WOD} \\
\frac{d\Q}{dt}&\equiv- \frac{\partial U}{\partial x} \dot x - \frac{d \K}{dt} \label{Q}
\end{align}
We show in Refs.~\cite{Dago-2022-PRL,Dago-2022-JStat} that the average heat flow is given by 
\begin{align} \label{QQ}
\frac{d\langle\Q\rangle}{dt}&=\frac{\omega_0}{Q} (2 \langle \K \rangle-k_B T_0).
\end{align}
Let us point out that this last expression is completely general and doesn't depend on the potential shape or current transformations occurring in the system. It also highlights that for a large quality factor $Q$, the heat exchanges with the thermal bath are reduced. Finally, at equilibrium, the equipartition theorem (Eq.~\ref{Eqpart_Ec}) implies that there are in average no heat exchanges, as expected.

\section{Kinetic temperature}
\label{supp_temp}
We define the kinetic temperature $T$ of the first deflection mode of the system through the velocity variance $\sigma_v^2=\langle v^2 \rangle-\langle v \rangle^2$:
\begin{align}
T=\frac{m}{k_B} \sigma_v^2 .\label{kinetic_temp}
\end{align}

The above definition can be reframed using the average kinetic energy $\langle K \rangle =\frac{1}{2} m \langle v^2 \rangle $, after introducing the deterministic kinetic energy contribution, $K_D(t)=\frac{1}{2} m \langle v \rangle ^2=\frac{1}{2} m \dot x_D^2$ (with $x_D(t)$ the solution of the deterministic equation of motion~\cite{Dago-2022-PRL,Gomez}):
\begin{align}
T(t)=\frac{2}{k_B} \big[\langle K(t) \rangle-K_D(t)\big],
\end{align}
where we explicitly indicated the possible time dependence of all quantities: the temperature can evolve in time depending on the power balance between external work and heat exchanges. Let us point out here that the kinetic temperature is only well defined as long as the speed PDF remains gaussian and therefore described by its variance $\sigma_v^2$. We observe using numerical simulations that the above assumption becomes an approximation for very fast procedures ($\mathbf v_1 >0.5$) and is no longer valid for $\mathbf v_1 >1$. 

Plugging in \eqref{QQ} the kinetic temperature expression, the average heat can be expressed as: 
\begin{align}
\frac{d\langle \Q(t)\rangle}{dt}&=\frac{\omega_0}{Q}\big(m\langle v(t) \rangle^2 + k_B (T(t) - T_0)\big). 
\label{Qgen}
\end{align}
At equilibrium in a potential that does not depends on the velocity, the kinetic temperature should match the bath temperature $T_0$ as prescribed by the equipartition [\eqref{Eqpart_Ec}]. Besides, when the deterministic terms are negligible compared to the thermal ones, $ \langle v \rangle ^2 \ll \sigma_v^2$, the average kinetic energy is proportional to the kinetic temperature, $\langle \K \rangle = \frac{1}{2}k_B T$, and the average heat simplifies into:
 \begin{align}
\frac{d\langle \Q(t) \rangle}{dt}=\frac{\omega_0}{Q}k_B (T(t)-T_0)
\label{meanQ_T}
\end{align}

For large quality factors, heat exchanges with the bath are negligible: $d \langle \Q \rangle =0$. This is the adiabatic regime.

\section{Adiabatic limit using phase space volume conservation}\label{sec_phase_volume}

\begin{figure}[!b]
\begin{center}
	\includegraphics[width=7cm]{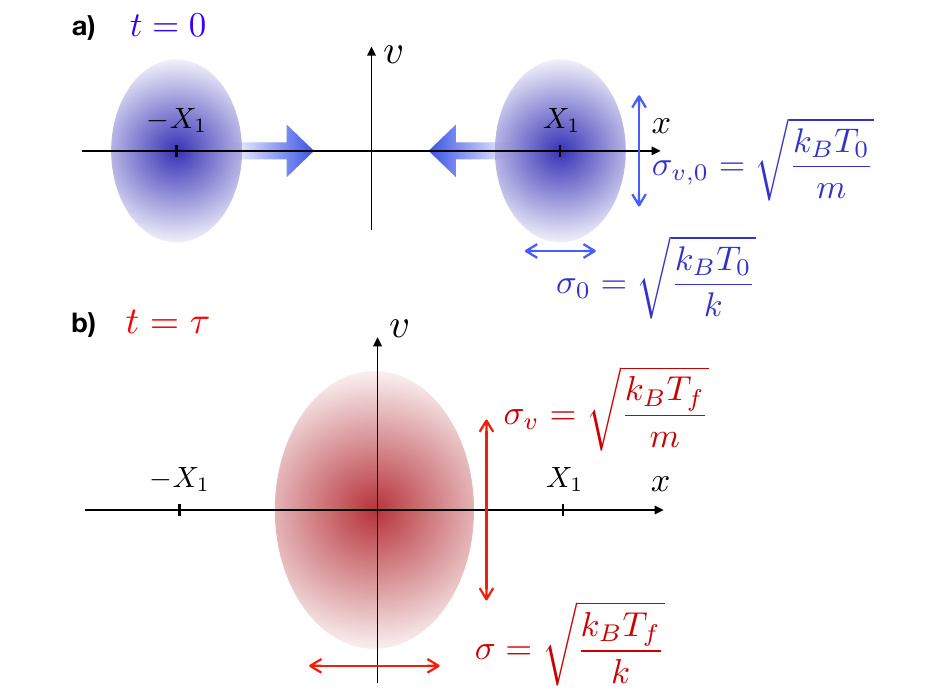}
\end{center}
	\caption{ \textbf{Schematic view of the phase space evolution during an adiabatic compression. (a) Initial time $t=0$ before the adiabatic compression}. When $X_1\gg\sigma_0$, the initial available phase space corresponds to two 2D not overlapping Gaussians centered in $\pm X_1$ and of variances $\sigma_0=\sqrt{k_BT_0/k}$ in position and $\sigma_{v,0}=\sqrt{k_BT_0/m} $ in velocity. The latters correspond to the equilibrium ones in the bath at temperature $T_0$. The systems has the same phase volume as the combination of two distinct harmonic oscillators, expressed in \eqref{J0}. \textbf{(b) Final time $t=\tau$, at the end of the adiabatic compression.} The phase space volume has evolved into the red ellipse, expressed in \eqref{Jf}, corresponding to a single harmonic oscillator at temperature $T_f$: a 2D Gaussian centered in 0 of variances $\sigma=\sqrt{k_BT_f/k}$ and $\sigma_{v}=\sqrt{k_BT_f/m}$ (see PDF expression in section~\ref{PDF}). In the adiabatic limit, the conservation of the phase space volume results in \eqref{conservation_temp}: $T_f=T_a=2T_0$.} \label{Liouville}
\end{figure}

The adiabatic limit case can be described easily without going through the whole model complex solving: we propose here an alternative demonstration to the one in the Supp.~Mat. of Ref.~\citenum{Dago-2022-PRL}. The highest temperature rise happens for fast erasures at high $Q$, when the heat exchanges with the bath are negligible: $d \langle \Q \rangle =0$. Such situation corresponds to protocols much faster than the relaxation time, {\it i.e.} when $\tau \ll \tau_\mathrm{relax} = 2 Q/\omega_0$. In this case, step 1 [Merge] corresponds to an adiabatic compression (or mean adiabatic~\cite{martinez_adiabatic_2015}) in the sense that the phase space is compressed without heat exchanges with the environment. In these conditions, the system is isolated from the bath and its evolution can be considered Hamiltonian. The phase space density is therefore conserved during the transformation according to Liouville's theorem~\cite{Peliti_book}. 

In other words, the phase space volume enclosed by the energy surface defined by the system's energy at every moment $E(t)$~\cite{martinez_adiabatic_2015} is conserved. For transformations slow enough, {\it i.e.} for $\mathbf v_1 \ll 1$, such volume will conserve a well defined shape during its time evolution (formally, the adiabatic invariants are preserved under the adiabatic principle~\cite{Tong_book,Neishtadt_2019}). Fig.~\ref{Liouville} illustrates the change in the phase space volume during step 1 [Merge] performed in an adiabatic fashion (high quality factor).
The system's energy corresponds to the Hamiltonian for a driving parameter $x_1$, $H_{x_1}$ expressed as $H_{x_1}(x,v)=\frac{1}{2}k(\lvert x \rvert -x_1)^2+\frac{1}{2}mv^2$. In the following we call $J(E,x_1)$ the phase space volume enclosed by the average energy $E$:
\begin{align}
	J(E,x_1)=\int_{H_{x_1}<E}dvdx
	\label{defJ}
\end{align}
At the initial time of the adiabatic compression, we have $x_1=X_1\gg\sigma_0$. Therefore, we can separate the $x>0$ and the $x<0$ in two elliptic volumes that do not overlap (see Fig.~\ref{Liouville}). We introduce $H_{x_1}^\pm=\frac{1}{2}k(x\mp x_1)^2+\frac{1}{2}mv^2$ corresponding respectively to the positive and negative x phase spaces. We can express the initial phase space volume $J_i(E)=J(E,X_1)$ [in blue in Fig.~\ref{Liouville}(a)] using \eqref{defJ}: 
\begin{align}
	J_i(E)&=\int_{H_{X_1}<E}dvdx \\
	&=\int_{H_{X_1}^+<E}dvdx + \int_{H_{X_1}^-<E}dvdx \\
	&=2\int_{H_{X_1}^+<E}dvdx \\
	&=\frac{4\pi E}{\sqrt{mk}} \label{J0}
\end{align}
\eqref{J0} derives from the area $\pi ab$ of the phase space delimited by the elliptic equation $(x-x_1)^2/a^2+v^2/b^2=1$ with $a=\sqrt{2E/k}$ and $b=\sqrt{2E/m}$.

Similarly, at the final time of the adiabatic compression we have $x_1=0$, and therefore we can express the final phase volume $J_f(E)=J(E,0)$ (in red in Fig.~\ref{Liouville}b): 
\begin{align}
	J_f(E)&=\int_{H_{0}<E}dvdx \\
	&=\frac{2\pi E}{\sqrt{mk}} \label{Jf}
\end{align}
Assuming the elliptic shape in the phase space for the final state supposes that it is correctly described by a Boltzmann statistics, which holds for slow enough protocols ($\mathbf v_1 \ll 1$, in practice $\mathbf v_1 < 0.5$ is sufficient).

Using the conservation of the phase space mentioned above, and knowing that the mass and stiffness are the same in \eqref{J0} and \eqref{Jf}, we have:
\begin{align}
	J_i(E_i)&=J_f(E_f)\\
	\Rightarrow E_f&=2E_i \label{conservation_energie}
\end{align}
As for $X_1\gg \sigma_0$ the system starts and ends in a quadratic potential, the average energy in the initial and final states satisfies $\langle E_i \rangle /T_0=\langle E_f \rangle /T_f=k_B$ [see \eqref{equipartcomp}]. The latter derived in section~\ref{PDF} relies on the gaussian ansatz of the speed and position PDF valid for $\mathbf v_1 \lesssim 0.5$. That is why we deduce straightforwardly from \eqref{conservation_energie} that:
\begin{align}
	T_f=2T_0 \label{conservation_temp}
\end{align}
Hence, the adiabatic limit in temperature is $T_a=2T_0$. As a consequence, the average work required in the adiabatic limit is given by the first law of thermodynamics (with $\langle Q \rangle =0$, see section~\ref{supp_thermo} for details):
\begin{align}
	\langle \W_S \rangle&= \Delta \langle E \rangle\\
	&=k_BT_f-k_BT_0\\
	&=k_BT_0=\W_a
	\label{eq_Wa}
\end{align}

During this demonstration, we used two hypotheses: adiabatic means fast, so that $\tau \ll \tau_\mathrm{relax} = 2 Q/\omega_0$, and but the protocol should at the same time be slow, so that $\mathbf{v}_1\ll 1$, ie $\tau \gg \mathbf{X}_1/\omega_0$. Any system and protocol allowing both criteria to coexist can be considered in the adiabatic and quasistatic regime. Such situation requires $2 Q \gg \mathbf{X}_1$, where $\mathbf{X}_1$ can be expressed in terms of barrier height $\mathcal{B}$ for a generic double well potential: $\mathbf{X}_1\sim \sqrt{2 \mathcal{B} / (k_B T_0)}$. In our case, with $\mathcal{B}\sim 12 k_B T_0$ and $Q\sim 80$, we have a reasonable window to demonstrate the adiabatic limit. From a pragmatic point of view, we can note a posteriori that the model works well for a range of parameters in our set-up, and what are the effective limit of applicability of the adiabatic regime.

\section{Probability Distribution Function during the compression}
\label{PDF}

To compute the other energetic terms ($\langle K \rangle$ and $\langle U \rangle$) during stage 1, we rely on the PDF of position $x$ and speed $v$. Let us introduce this PDF during the compression stage, supposing that the system is at equilibrium: it is governed by the Boltzmann distribution
\begin{align}
P^{c}(x,v) &= \frac{1}{Z^{c}} e^{-\frac{1}{2} \beta m v ^2} e^{-\frac{1}{2} \beta k (| x | -x_1)^2}, \label{suppPc}
\end{align}
with $\beta=1/ (k_B T)$ and $Z^{c}$ the partition function:
\begin{align}
Z^{c}(\beta,x_1) &= \frac{2 \pi}{\sqrt{km} \beta}\V, \label{suppZc}
\end{align}
where $\V$ is a volume-like function that shrinks by a factor 2 when $x_1$ decreases from $X_1$ to $0$:
\begin{align}
\V(\beta,x_1)&=1+\erf\left(\sqrt{\frac{k\beta}{2}} x_1\right). \label{suppV}
\end{align}
We can directly apply this PDF to the slow erasures, in equilibrium at temperature $T_0$ at all time. We extend the use of this PDF to the case of fast compression as well, under the hypotheses that (i) the cantilever oscillates several times in the double-well before its shape changes significantly ($|\mathbf v_1|<0.5$), so that the phase space is adequately sampled and (ii) a Boltzmann-like distribution still holds. In this case, however, we let the temperature $T$ as a parameter free to evolve due to a possible heating. Note that the PDF $P^c(x,v)$ only describes the volume compression and does not include any transients, leaving aside any coupling between $x$ and $v$. The main transient during the basic protocol, due to the sudden translational motion of the wells, is addressed in Ref.~\citenum{Dago-2022-PRL}. The relevancy of the ansatz (for $\mathbf v_1 \lesssim 1$) is also demonstrated in Ref.~\citenum{Dago-2022-PRL}. This transient essentially disappear using an optimal translation protocol.

In the initial and final states of the compression, assuming that $x_1(0)=X_1\gg\sigma_0$, the double potential behaves as a single harmonic one and therefore the PDF in \eqref{suppPc} simplifies respectively into (indices $i$ and $f$):

\begin{subequations} \label{eq.P_simp}
\begin{align}
P^{c}_i(x,v) &= \frac{1}{Z^{c}} e^{-\frac{1}{2} \beta_i m v ^2} e^{-\frac{1}{2} \beta_i k (x-X_1)^2} \label{Pci}\\
P^{c}_f(x,v) &= \frac{1}{Z^{c}} e^{-\frac{1}{2} \beta_f m v ^2} e^{-\frac{1}{2} \beta_f k x^2} \label{Pcf}
\end{align}\end{subequations}

 Hence, the average energy in the initial and final states of the compression writes: 
 \begin{align}
\langle E_{i,f} \rangle &=\iint \left( \frac{1}{2}mv^2+\frac{1}{2}k\big[x-x_1(t_{i,f})\big]^2\right)P^{c}_{i,f}(x,v) dx dv \nonumber\\
&=\frac{1}{\beta_{i,f}} =k_B T_{i,f} \label{equipartcomp}
\end{align}

\eqref{equipartcomp} reads as equipartition at the memory's temperature $T(t)$ (possibly different from $T_0$): it results from the Boltzmann like PDF approximation valid as long as the driving speed is not too high.
\section{Optimal protocol}
\label{supp_optimal}

 The optimal erasure protocol consists in adapting the translational driving to always minimize the viscous work. In other words we replace all the linear translational drivings by optimal protocols explicitly computed in Ref.~\citenum{Gomez} (and also tackled in Ref.~\citenum{Dellago}). Basically, the deterministic translational motion is now done at an optimal constant speed, getting rid of the transient oscillations, by applying force peaks at the initial and final instant of the ramp as displayed in dashed red on Fig.~\ref{meanz}b. The analytic expression of the optimal translational driving between position $0$ and $-X1$ in a duration $\tau$ is~\cite{Gomez}:
\begin{align}
x^{\textrm{opt}}_1(t)=-X_1 \frac{kt/\gamma+1}{k\tau/\gamma+2}-\frac{mX_1}{2\gamma+k\tau} [\delta(t)-\delta(t-\tau)],
\end{align}
with $k$ the cantilever stiffness, $m$ its mass and $\gamma$ the damping coefficient.

This driving is applied as such during Step2 [Translate] using the potential $U_2^{\textrm{opt}}(x,t)= \frac{1}{2}k(x -x^{\textrm{opt}}_1(t))^2$. During Step1 [Merge] the optimal driving is also applied symmetrically on the two wells using the potential $U_1^{\textrm{opt}}(x,t)= \frac{1}{2}k(| x | -x^{\textrm{opt}}_1(t))^2$: the optimal initial and final kicks are computed likewise for the left and right wells. The full optimal erasure protocol is plotted on Fig.~\ref{intro}c).

\section{Power profile evolution during an optimal erasure}

\begin{figure}[htb]
\begin{center}
	\includegraphics[width=8.4cm]{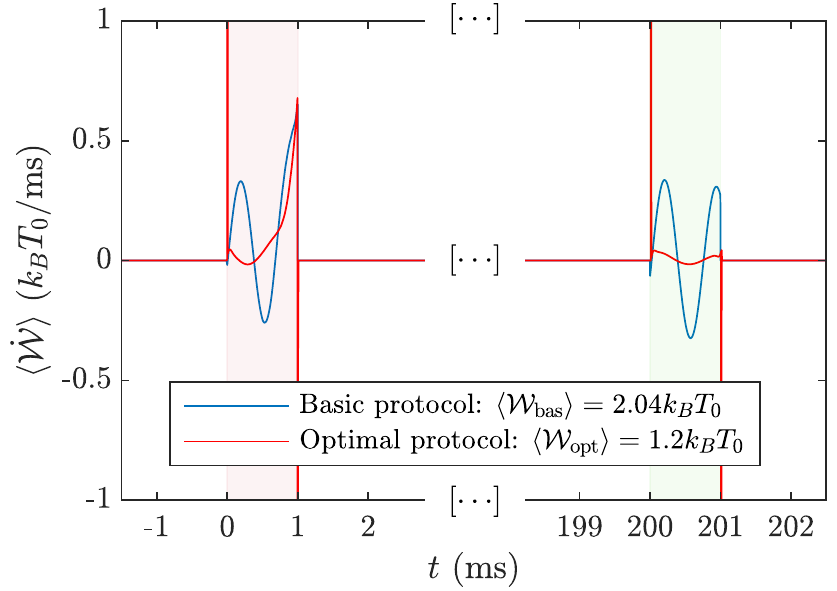}
\end{center}
	\caption{\textbf{Mean power measured during an erasure protocol.} Optimal protocols (red) demonstrates their efficiency compared to basic protocols (blue): the optimal erasure suppresses the oscillation in the average work due to the translational motion deterministic transient in the basic approach. The positive and negative peaks in $\langle \dot{\W} \rangle$ induced by the driving pulses compensate each-other and do not impact the total work. The overall cost is reduced from $\langle \W_\mathrm{bas} \rangle =(2.04 \pm 0.05 )k_BT_0$ to $\langle \W_\mathrm{opt} \rangle= (1.20\pm 0.04 )k_BT_0$.} \label{dW}
\end{figure}

In addition to Fig.~\ref{Approach_vide_W2} comparing the erasure cost between basic and optimal erasure protocols we propose here to study into further details the mean power required during the procedure. We show in Fig.~\ref{dW} the mean power evolution for an erasure of duration $\tau=\SI{1}{ms}$ (corresponding to the fastest experimental data). In blue the response to the basic protocol emphasizes the dissipation contribution visible through the transient oscillations. In particular, when the steps durations match half the oscillation period (modulo one period) as it is the case here, the dissipation work culminates. This can be understood by the fact that the positive area under the last half oscillation is not compensated by a negative one. This is the reason why the dissipation cost oscillates with the erasure speed as visible in Fig.~\ref{Approach_vide_W2}. On the contrary the optimal response is expected to suppress the mean power oscillations during step 2 translation as it makes the system position sticks to the driving, in agreement with Fig.~\ref{meanz}. We see on Fig.~\ref{dW} that not only the step 2 transient oscillations are deleted, but also the ones in step 1. This observation demonstrates the efficiency of the optimal translational driving, even in the double-well potential. All in all the overall cost is almost reduced by half using the optimal procedure. Hence the optimal procedure minimizes the erasure cost and removes the oscillations in the $\langle \W \rangle$ vs $\mathbf v _1$ evolution. Finally, one might also wonder about the peaks on the red curve: it corresponds to the force impulse imposed to the system to initiate the translation and to the work recovered when suddenly slowing down the system with the second dirac force and the end.

\section{Model extended to optimal protocol: average erasure cost map}
\label{supp_model}

\begin{figure}[htb]
	\includegraphics[width=\columnwidth]{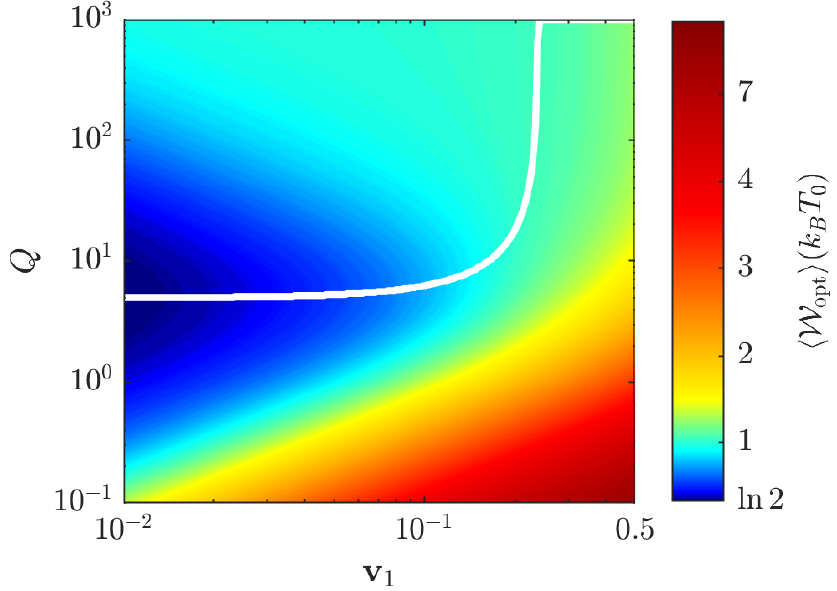}
	\caption{\textbf{Optimal erasure cost map} obtained using the model adapted to the optimal protocol. The average work required for an optimal erasure is coded by the color scale on the right, and plotted as a function of the protocol speed $\mathbf v_1$ and the memory quality factor $Q$. The white curve corresponds to the quality factor minimizing the average erasure cost at any given speed. On the first hand, optimizing very fast erasures ($\mathbf v_1>0.1$) means choosing high quality factors. On the other hand, for moderate speeds ($\mathbf v_1<0.1$) the optimal damping is around $Q=5$.}
	\label{Map}
\end{figure}

We extend the model detailed in Ref.~\citenum{Dago-2022-PRL} to the optimal erasure protocol by 1) replacing the deterministic contribution $\W_D$ by the optimal translational cost $\W_{D,\textrm{opt}}$ expressed in \eqref{W2opt}; and 2) adding the kinetic energy given to the system through the initial force peak at the beginning of step 1, and assumed to not be recovered through the final force peak at the final instant. These modifications of the model perfectly describe step 2, and approximate well the system's response during step 1: we illustrate in Fig.~\ref{intro} its good agreement with the simulation results. The slight overestimation of the optimal erasure cost observed on Fig.~\ref{intro} is likely to come from the fact that the final force peak is not totally inefficient to bring back the average velocity to 0 after step 1 as assumed in the model. Furthermore, at high speeds ($\mathbf v_1 > 0.5$) the model assumption stops being true, causing the deviation from the simulation results. Despite these small approximations, the model is reliable enough to estimate the optimal erasure cost map in Fig.~\ref{Map}. We deduce from it that for moderate speeds ($\mathbf v_1 <0.1$) the optimal quality factor is $Q\sim5$, while when the erasure speed increases one should rather operate without any dissipation ($Q\to \infty$). The above conclusion about the optimal quality factor drawn from the erasure cost map is consistent with the analysis based on the balance between the average stochastic warming cost and deterministic dissipation one.}

\showmatmethods{} 

\acknow{This work has been financially supported by the Agence Nationale de la Recherche through grant ANR-18-CE30-0013 and by the FQXi Foundation, Grant No. FQXi-IAF19-05, “Information as a fuel in colloids and superconducting quantum circuits.” We thank J. Pereda for fruitful scientific discussions and for the initial FPGA programming.}

\showacknow{} 

\bibliography{AdiabaticComputing}

\end{document}